\documentclass[a4paper,11pt]{article}
\usepackage{pos}

\usepackage{xspace}
\def\sss{\scriptscriptstyle}
\newcommand{\as}{\alpha_{\sss\rm S}}
\newcommand{\aew}{\alpha_{\sss\rm EW}}
\newcommand{\pt}{p_{t}}
\newcommand{\noun}[1]{\textsc{#1}}

\newcommand{\NLOPS}{NLO\textsubscript{QCD}+PS\xspace}
\newcommand{\NLOEWPS}{NLO\textsubscript{EW}+PS\xspace}
\newcommand{\NLONLOPS}{NLO\textsubscript{QCD}+NLO\textsubscript{EW}+PS\xspace}
\newcommand{\NNLONLOPS}{NNLO\textsubscript{QCD}+NLO\textsubscript{EW}+PS\xspace}
\newcommand{\NNLOPS}{NNLO\textsubscript{QCD}+PS\xspace}
\newcommand{\minloprime}{\noun{MiNLO$^\prime$}\xspace}
\newcommand{\geneva}{\noun{Geneva}\xspace}
\newcommand{\unnlops}{\noun{Unnlops}\xspace}
\newcommand{\vincia}{\noun{Vincia}\xspace}
\newcommand{\powheg}{{\ttfamily POWHEG}\xspace}
\newcommand{\minnlops}{\textsc{MiNNLO\textsubscript{PS}}\xspace}

\title{Recent progress on high order calculations and matching to parton showers}

\author*[a,1]{Emanuele Re}

\affiliation[a]{LAPTh, Universit\'e Grenoble Alpes, Universit\'e Savoie Mont Blanc, CNRS,\\
  F-74940 Annecy, France}

\note{Preprint number: LAPTH-Conf-029/21. Slightly extended version of
  the one submitted for the proceedings.}

\emailAdd{emanuele.re@lapth.cnrs.fr}

\abstract{I give an overview of the recent progress on the matching
  of fixed-order calculations and parton showers. The focus is on the
  matching with NNLO QCD corrections as well as with NLO EW ones.}

\FullConference{%
  
 The Ninth Annual Conference on Large Hadron Collider Physics - LHCP2021\\
 7-12 June 2021\\
 Online
}

\begin{document}
\maketitle

\section{Introduction}

The current experimental precision reached by the LHC experimental
collaborations (and even more the future prospects) requires
theoretical predictions whose formal accuracy goes beyond the
computation of NLO QCD corrections and their matching to parton
showers (\NLOPS). Including (N)NNLO QCD corrections, and NLO EW ones (or
combination thereof), is crucial, as often it is only through such
predictions that a comparison between data and theory is made possible
without being limited by the quality of theoretical predictions.

It is known that, in the corners of phase space where an hierarchy
between two or more scales develops, fixed-order QCD predictions fail,
due to the presence of large logarithms, that need to be resummed to
all orders. The recent progress in this field has been remarkable as
well, not only due to the extremely precise predictions obtained for
specific observables,
but also because of the
development of parton-shower (PS) algorithms whose logarithmic accuracy can
be formally established for a wide class of observables, and
systematically improved.

A fully-differential (parton-shower) Monte Carlo event generator that incorporates,
consistently, several of the above developments does not exist yet. In
the rest of this review, I'll summarize the recent progress in the
matching of NNLO QCD computations with parton showers (\NNLOPS), and
in the inclusion of EW corrections in event generators.

\section{Matching NNLO QCD corrections with Parton Showers}
The issue of matching NNLO QCD corrections to PS has been already
addressed by different groups, and \NNLOPS results have been obtained
with four methods: ``reweighted
\minloprime''~\cite{Hamilton:2012rf,Hamilton:2013fea},
\geneva~\cite{Alioli:2012fc,Alioli:2013hqa},
\unnlops~\cite{Hoeche:2014aia},
\minnlops~\cite{Monni:2019whf,Monni:2020nks}.\footnote{After this talk
  was given, proof-of-concepts results using another \NNLOPS method,
  based on the separation of strongly-ordered and unordered regions as
  obtained, in \vincia, through so-called ``sector showers'', have
  been presented in Ref.~\cite{Campbell:2021svd}.}
Very schematically, the core ideas of these four methods can be summarized as follows:
\begin{itemize}
\item ``reweighted'' \minloprime and \minnlops are based on the merging of \NLOPS
  results for $pp\to F$ and $pp\to F+j$ production, where $F$ denotes
  a generic color-singlet final state. Such merging is obtained
  without any external resolution parameter, through the use of
  information known from transverse-momentum resummation. The extra
  terms needed to get NNLO accuracy for inclusive observables have
  been included, in the older implementations, through a numerical
  ``reweighting'' of the events, and, more recently, through a
  novel formulation of the method (\minnlops, ultimately based on the
  $\pt$-resummation method proposed in
  Ref.~\cite{Monni:2016ktx,Bizon:2017rah}) where the NNLO corrections
  are included analytically from scratch.
\item In \geneva, one constructs IR-finite events using a resolution
  parameter (until recently, the ``$N$-jettiness'' $\tau_N$) whose
  resummation properties are accurately known, and that, through a cut
  ($\tau^{\rm cut}_N$), allows one to translate an ``$M$-parton''
  event to an ``$N$-jet'' event. The extra radiation is provided by a
  parton shower, which, however, needs to be constrained by a
  requirement on $\tau^{\rm cut}_N$.
\item In \unnlops, one first promotes to NLO accuracy an
  ``unitarized'' CKKW approach, by carefully adding higher order
  contributions, and removing the pre-existing approximate terms at
  order $\as$. The missing NNLO ingredients are then supplemented
  subsequently.
\end{itemize}
All the processes with 2 massless colored legs at LO can be described
with \NNLOPS accuracy, and many results have been already
obtained~\cite{Hamilton:2013fea,Karlberg:2014qua,Astill:2016hpa,Astill:2018ivh,Re:2018vac,Bizon:2019tfo,Alioli:2015toa,Alioli:2019qzz,Alioli:2020fzf,Alioli:2020qrd,Alioli:2021qbf,Alioli:2021egp,Cridge:2021hfr,Hoeche:2014aia,Hoche:2014dla,Hoche:2018gti,Monni:2019whf,Monni:2020nks,Lombardi:2020wju,Lombardi:2021rvg,Buonocore:2021fnj,Hu:2021rkt},
including, in one case (based on the \minloprime idea), results where
NLO QCD accuracy was retained not only for the first jet, but also for
the second one~\cite{Frederix:2015fyz}.

\begin{figure}[!htb]
  \begin{minipage}{0.5\linewidth}
    \centerline{\includegraphics[scale=0.4]{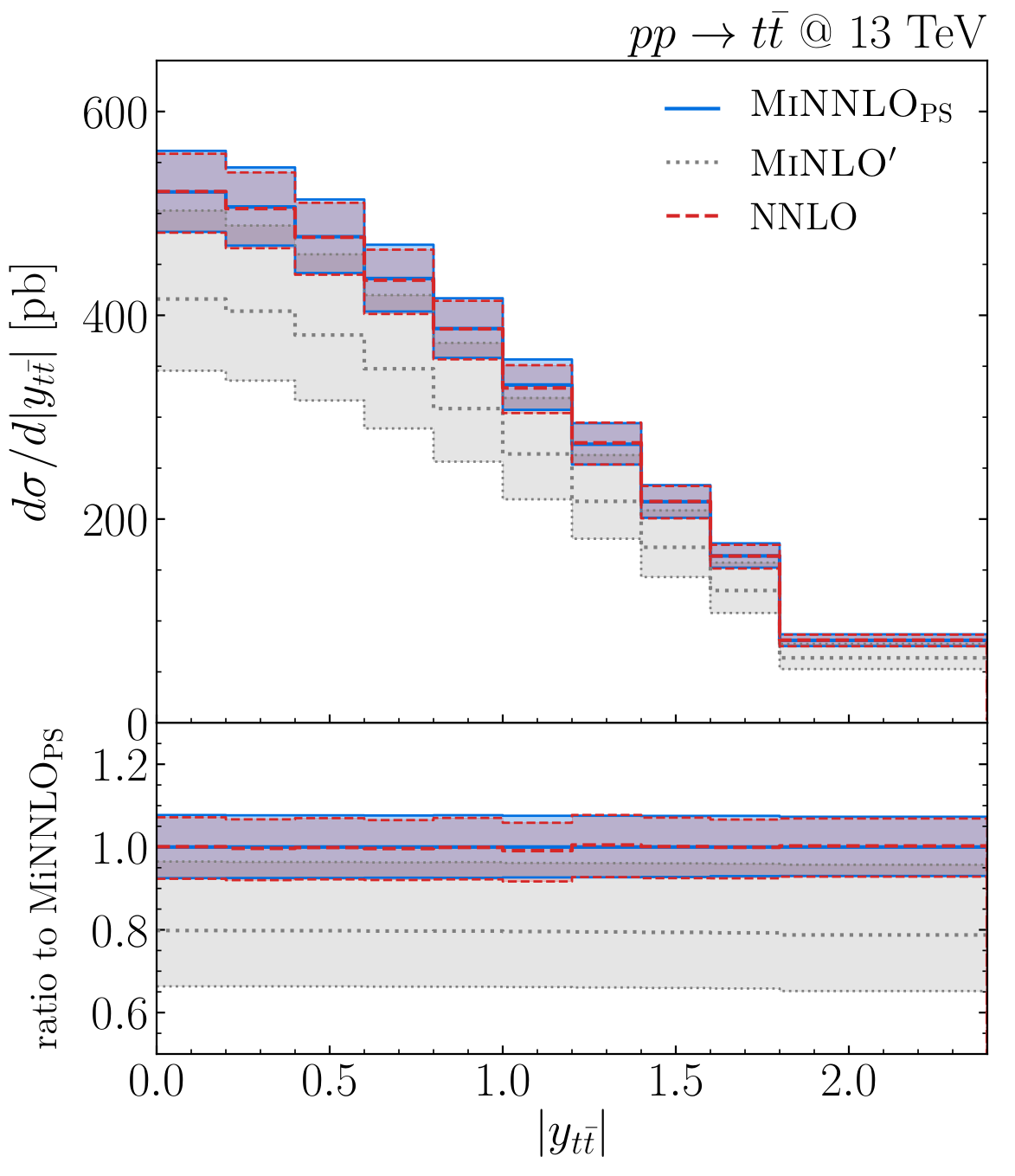}}
  \end{minipage}
  \begin{minipage}{0.5\linewidth}
    \centerline{\includegraphics[scale=1.3]{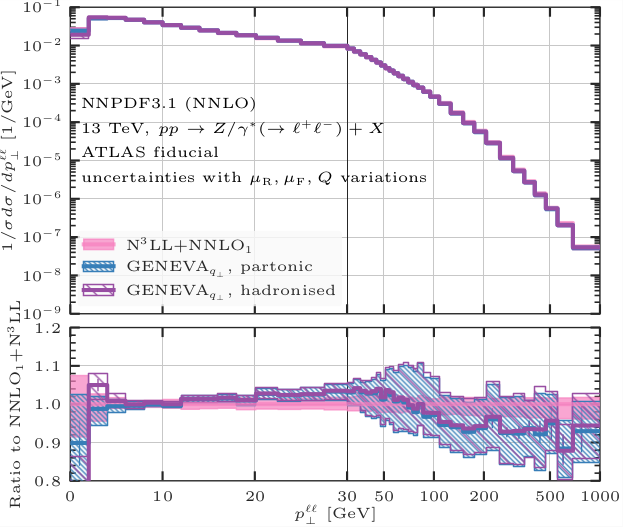}}
  \end{minipage}
  \caption{Left (supplemental material of
    Ref.~\cite{Mazzitelli:2020jio}): comparison of NNLO (red),
    \minloprime (gray), and \minnlops (blue) results for the
    $t\bar{t}$ rapidity in top-pair production. Right
    (Ref.~\cite{Alioli:2021qbf}): comparison of the $\pt$ spectrum of
    the $Z$ boson in Drell-Yan production as obtained with \geneva
    (blue and violet) and with {\tt Radish}+{\tt NNLOJET} (pink).}
  \label{fig:QCDplots}
\end{figure}

Besides the huge number of results for color-singlet production, in
the last few months the first-ever \NNLOPS results for a process
beyond color-singlet production were obtained: in
Ref.~\cite{Mazzitelli:2020jio} NNLO QCD corrections were matched to
parton showers for top-pair production, through a non-trivial
extension of the \minnlops method. Recent progress with the \geneva
method notably includes the first results obtained with a resolution
parameter for the ``0 to 1 jet'' region other than
$\tau_0$~\cite{Alioli:2021qbf}, where {\tt Radish}
results~\cite{Monni:2016ktx,Bizon:2017rah,Bizon:2018foh} for $\pt$
resummation at N3LL were used as an input. In
Ref.~\cite{Prestel:2021vww} the \unnlops method was generalized to
take into account N3LO QCD corrections. Fig.~\ref{fig:QCDplots} shows
results for $t\bar{t}$ production (with \minnlops) and for the $\pt$
spectrum of the $Z$ boson in Drell-Yan production (with \geneva).

\section{Matching NLO QCD and NLO EW corrections with Parton Showers}
The computation of NLO EW corrections to multileg processes at the LHC
can be considered a conceptually solved problem. In practice, loop
corrections at order $\aew$ can be obtained through the use of
automated 1-loop providers, and full NLO results can be obtained
combining virtual and tree-level real corrections, through subtraction
methods.

Building up from \NLOPS methods, it is possible to match NLO EW
corrections with parton showers (\NLOEWPS), using different approaches
that also allow, at least for simple processes, a simultaneous
matching of QCD and EW corrections. Nevertheless, it is certainly true
that a fully-general approach to tackle this challenge, and that is
valid also for processes featuring QCD/EW interference at LO (as, for
instance, $pp\to t\bar{t}$) is still missing, and it is one of the
open problems in the field.

\begin{figure}[!htb]
  \begin{minipage}{0.5\linewidth}
    \centerline{\includegraphics[scale=0.9]{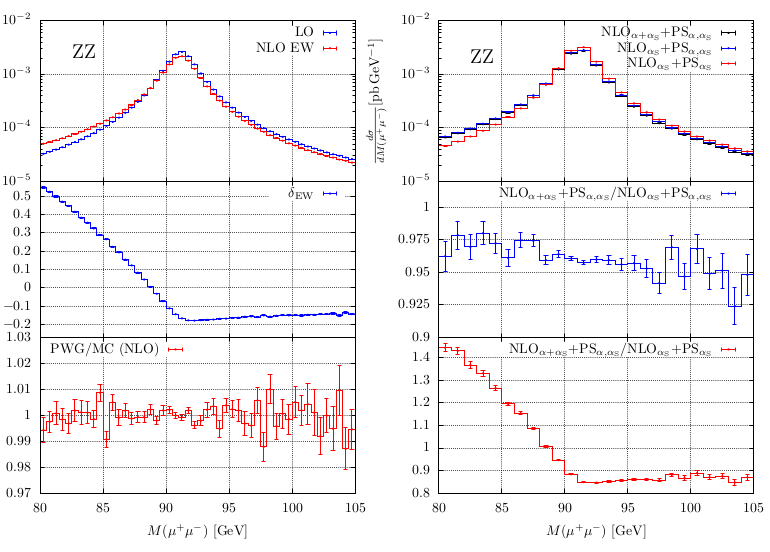}}
  \end{minipage}
  \begin{minipage}{0.5\linewidth}
    \centerline{\includegraphics[scale=0.9]{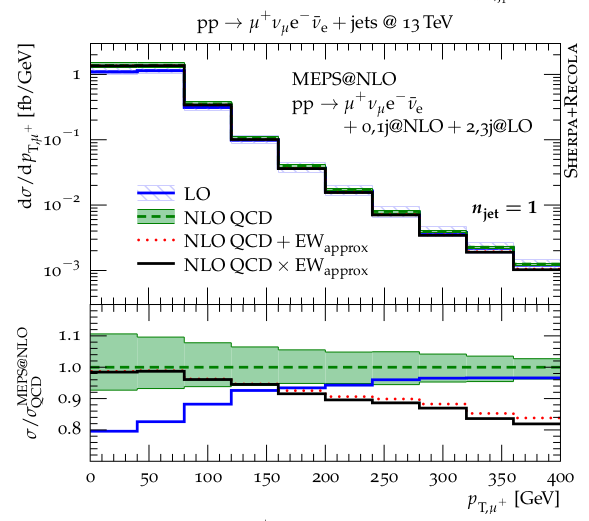}}
  \end{minipage}
  \caption{Left (Ref.~\cite{Chiesa:2020ttl}): Results, at different
    orders and approximations, for the invariant mass of the muonic
    pair in $pp\to e^+e^- \mu^+\mu^-$ production. Right
    (Ref.~\cite{Brauer:2020kfv}): Results with and without EW effects
    for the muon transverse momentum in NLO merged predictions for
    $pp\to \mu^+\nu_{\mu} e^- \bar{\nu}_{e} +$ jets.}
  \label{fig:EWplots}
\end{figure}

There are currently two approaches to get \NLONLOPS results:
\begin{itemize}
\item One can include EW corrections through a local $K$-factor (which
  relies on the use of approximated integrated real contribution, and
  that acts on the ``Born'' configurations only) and by adding real
  QED radiation only through the parton shower. The main limitation of
  this scheme, at times denoted as {\tt EW\textsubscript{VI}} or {\tt
    EWvirt}, and first proposed in Ref.~\cite{Kallweit:2015dum}, is
  that, formally, it is not valid for hard photon radiation. It has
  been successfully used, though, for several
  processes~\cite{Kallweit:2015dum,Kallweit:2017khh,Gutschow:2018tuk,Czakon:2019bcq,Brauer:2020kfv},
  and, notably, to incorporate approximate electroweak corrections in
  \NLOPS merged
  simulations~\cite{Gutschow:2018tuk,Czakon:2019bcq,Brauer:2020kfv}.
\item An exact matching of the EW corrections (both of virtual and
  real origin) can be obtained using the {\tt POWHEG-BOX-RES}
  framework: this allows to include QCD and EW effects essentially
  through the traditional \powheg approach, but also allowing for the
  generation of strong or electromagnetic real radiation from each
  resonance simultaneously. QCD and EW corrections are combined
  exactly (additively) at order $\as$ and $\aew$, whereas factorizable
  and mixed $\as^n \aew^m$ terms are only included in the collinear
  limit. Such approach has been used in
  Refs.~\cite{Granata:2017iod,Chiesa:2019ulk,Chiesa:2020ttl} and,
  previously, for Drell-Yan production, in
  Refs.~\cite{Bernaciak:2012hj,Barze:2012tt,Barze:2013fru,Muck:2016pko,CarloniCalame:2016ouw,Chiesa:2019nqb}.
\end{itemize}
Two recent applications of the above methods are related to diboson
production processes ($4\ell$ production). Merged parton-shower
predictions for $p\to WW$ and $pp\to WW +\ 1\ \mbox{jet}$ production, that include NLO QCD and
EW corrections (the latter in the {\tt EWvirt} approximation), were
presented by the authors of Ref.~\cite{Brauer:2020kfv}. In
Ref.~\cite{Chiesa:2020ttl}, QCD and exact EW corrections to all
4-lepton final states were instead matched to parton showers through
the refinements of the \powheg method as implemented in the {\tt
  POWHEG-BOX-RES} framework. Fig.~\ref{fig:EWplots} displays a couple
of representative results taken from Ref.~\cite{Chiesa:2020ttl} (left)
and~\cite{Brauer:2020kfv} (right).

Other schemes have been recently proposed to lift some of the
limitations of the {\tt EWvirt} approach: for instance,
in Ref.~\cite{Bothmann:2020sxm}, an {\tt EWsud} scheme has been introduced,
where LL and NLL EW corrections are included in the Sudakov
limit~\cite{Denner:2000jv}, thereby allowing corrections to all jet
multiplicities. In this context, even an {\tt hybrid} scheme is being
studied~\cite{hybrid_psr_21}.

\section{Conclusions}
In this talk I have presented recent results related to the matching
of high order computations with parton showers. There has been an
intense activity in the last couple of years, mainly in the context of
matching at NNLO (\NNLOPS) and matching QCD and EW corrections
simultaneously (\NLONLOPS).

Among the future challenges, certainly there is the \NNLOPS matching
for a process with jets at LO, as well as the establishment of general
method(s) to get \NLONLOPS (and, eventually, \NNLONLOPS) accuracy for
processes featuring QCD/EW interference at LO.

The focus of this review was on matching aspects. The improvement
of parton-shower algorithms has been, recently, a very active area in
the field. Such activity covers several directions, spanning from the
inclusion of EW effects in parton showers to the introduction of a new
generation of parton shower algorithms whose accuracy goes beyond the
leading logarithm. It is likely that the impact of such developments
(the last one in particular) will be significant in the future. I
refer to the talks given, for instance, at the workshop ``Taming the
accuracy of event generators''~\cite{taming}, for a recent overview of
this activity.


\begin{thebibliography}{99}


\bibitem{Hamilton:2012rf} 
  K.~Hamilton, P.~Nason, C.~Oleari and G.~Zanderighi,
  JHEP {\bf 1305}, 082 (2013)
  [arXiv:1212.4504].

\bibitem{Hamilton:2013fea} 
  K.~Hamilton, P.~Nason, E.~Re and G.~Zanderighi,
  JHEP {\bf 1310}, 222 (2013)
  [arXiv:1309.0017 [hep-ph]].
 

\bibitem{Alioli:2012fc}
S.~Alioli, C.~W.~Bauer, C.~J.~Berggren, A.~Hornig, F.~J.~Tackmann, C.~K.~Vermilion, J.~R.~Walsh and S.~Zuberi,
JHEP \textbf{09} (2013), 120
[arXiv:1211.7049 [hep-ph]].

\bibitem{Alioli:2013hqa}
S.~Alioli, C.~W.~Bauer, C.~Berggren, F.~J.~Tackmann, J.~R.~Walsh and S.~Zuberi,
JHEP \textbf{06} (2014), 089
[arXiv:1311.0286 [hep-ph]].

  
\bibitem{Hoeche:2014aia}
S.~H\"oche, Y.~Li and S.~Prestel,
Phys. Rev. D \textbf{91} (2015) no.7, 074015
[arXiv:1405.3607 [hep-ph]].


\bibitem{Monni:2019whf}
P.~F.~Monni, P.~Nason, E.~Re, M.~Wiesemann and G.~Zanderighi,
JHEP \textbf{05} (2020), 143
[arXiv:1908.06987 [hep-ph]].

\bibitem{Monni:2020nks}
P.~F.~Monni, E.~Re and M.~Wiesemann,
Eur. Phys. J. C \textbf{80} (2020) no.11, 1075
[arXiv:2006.04133 [hep-ph]].


\bibitem{Campbell:2021svd}
J.~M.~Campbell, S.~H\"oche, H.~T.~Li, C.~T.~Preuss and P.~Skands,
[arXiv:2108.07133 [hep-ph]].


\bibitem{Monni:2016ktx}
P.~F.~Monni, E.~Re and P.~Torrielli,
Phys. Rev. Lett. \textbf{116} (2016) no.24, 242001
[arXiv:1604.02191 [hep-ph]].

\bibitem{Bizon:2017rah}
W.~Bizon, P.~F.~Monni, E.~Re, L.~Rottoli and P.~Torrielli,
JHEP \textbf{02} (2018), 108
[arXiv:1705.09127 [hep-ph]].

\bibitem{Karlberg:2014qua}
A.~Karlberg, E.~Re and G.~Zanderighi,
JHEP \textbf{09} (2014), 134
doi:10.1007/JHEP09(2014)134


\bibitem{Astill:2016hpa}
W.~Astill, W.~Bizon, E.~Re and G.~Zanderighi,
JHEP \textbf{06} (2016), 154
[arXiv:1603.01620 [hep-ph]].


\bibitem{Astill:2018ivh}
W.~Astill, W.~Bizo\'n, E.~Re and G.~Zanderighi,
JHEP \textbf{11} (2018), 157
[arXiv:1804.08141 [hep-ph]].


\bibitem{Re:2018vac}
E.~Re, M.~Wiesemann and G.~Zanderighi,
JHEP \textbf{12} (2018), 121
[arXiv:1805.09857 [hep-ph]].


\bibitem{Bizon:2019tfo}
W.~Bizo\'n, E.~Re and G.~Zanderighi,
JHEP \textbf{06} (2020), 006
[arXiv:1912.09982 [hep-ph]].



\bibitem{Lombardi:2020wju}
D.~Lombardi, M.~Wiesemann and G.~Zanderighi,
[arXiv:2010.10478 [hep-ph]].


\bibitem{Lombardi:2021rvg}
D.~Lombardi, M.~Wiesemann and G.~Zanderighi,
[arXiv:2103.12077 [hep-ph]].

\bibitem{Buonocore:2021fnj}
L.~Buonocore, G.~Koole, D.~Lombardi, L.~Rottoli, M.~Wiesemann and G.~Zanderighi,
[arXiv:2108.05337 [hep-ph]].



\bibitem{Hoche:2014dla}
S.~H\"oche, Y.~Li and S.~Prestel,
Phys. Rev. D \textbf{90} (2014) no.5, 054011
[arXiv:1407.3773 [hep-ph]].


\bibitem{Hoche:2018gti}
S.~H\"oche, S.~Kuttimalai and Y.~Li,
Phys. Rev. D \textbf{98} (2018) no.11, 114013
[arXiv:1809.04192 [hep-ph]].



\bibitem{Alioli:2015toa}
S.~Alioli, C.~W.~Bauer, C.~Berggren, F.~J.~Tackmann and J.~R.~Walsh,
Phys. Rev. D \textbf{92} (2015) no.9, 094020
[arXiv:1508.01475 [hep-ph]].


\bibitem{Alioli:2019qzz}
S.~Alioli, A.~Broggio, S.~Kallweit, M.~A.~Lim and L.~Rottoli,
Phys. Rev. D \textbf{100} (2019) no.9, 096016
[arXiv:1909.02026 [hep-ph]].


\bibitem{Alioli:2020fzf}
S.~Alioli, A.~Broggio, A.~Gavardi, S.~Kallweit, M.~A.~Lim, R.~Nagar, D.~Napoletano and L.~Rottoli,
JHEP \textbf{04} (2021), 254
[arXiv:2009.13533 [hep-ph]].


\bibitem{Alioli:2020qrd}
S.~Alioli, A.~Broggio, A.~Gavardi, S.~Kallweit, M.~A.~Lim, R.~Nagar, D.~Napoletano and L.~Rottoli,
JHEP \textbf{04} (2021), 041
[arXiv:2010.10498 [hep-ph]].


\bibitem{Alioli:2021qbf}
S.~Alioli, A.~Broggio, A.~Gavardi, S.~Kallweit, M.~A.~Lim, R.~Nagar, D.~Napoletanog, C.~W.~Bauer and L.~Rottoli,
[arXiv:2102.08390 [hep-ph]].


\bibitem{Alioli:2021egp}
S.~Alioli, A.~Broggio, A.~Gavardi, S.~Kallweit, M.~A.~Lim, R.~Nagar and D.~Napoletano,
Phys. Lett. B \textbf{818} (2021), 136380
[arXiv:2103.01214 [hep-ph]].

\bibitem{Cridge:2021hfr}
T.~Cridge, M.~A.~Lim and R.~Nagar,
[arXiv:2105.13214 [hep-ph]].


\bibitem{Hu:2021rkt}
Y.~Hu, C.~Sun, X.~M.~Shen and J.~Gao,
[arXiv:2101.08916 [hep-ph]].

\bibitem{Frederix:2015fyz}
R.~Frederix and K.~Hamilton,
JHEP \textbf{05} (2016), 042
[arXiv:1512.02663 [hep-ph]].



\bibitem{Mazzitelli:2020jio}
J.~Mazzitelli, P.~F.~Monni, P.~Nason, E.~Re, M.~Wiesemann and G.~Zanderighi,
Phys. Rev. Lett. \textbf{127} (2021) no.6, 062001
[arXiv:2012.14267 [hep-ph]].


\bibitem{Bizon:2018foh}
W.~Bizo\'n, X.~Chen, A.~Gehrmann-De Ridder, T.~Gehrmann, N.~Glover, A.~Huss, P.~F.~Monni, E.~Re, L.~Rottoli and P.~Torrielli,
JHEP \textbf{12} (2018), 132
[arXiv:1805.05916 [hep-ph]].


\bibitem{Prestel:2021vww}
S.~Prestel,
[arXiv:2106.03206 [hep-ph]].







\bibitem{Kallweit:2015dum}
S.~Kallweit, J.~M.~Lindert, P.~Maierhofer, S.~Pozzorini and M.~Sch\"onherr,
JHEP \textbf{04} (2016), 021
[arXiv:1511.08692 [hep-ph]].

\bibitem{Gutschow:2018tuk}
C.~G\"utschow, J.~M.~Lindert and M.~Sch\"onherr,
Eur. Phys. J. C \textbf{78} (2018) no.4, 317
[arXiv:1803.00950 [hep-ph]].

\bibitem{Czakon:2019bcq}
M.~L.~Czakon, C.~G\"utschow, J.~M.~Lindert, A.~Mitov, D.~Pagani, A.~S.~Papanastasiou, M.~Sch\"onherr, I.~Tsinikos and M.~Zaro,
[arXiv:1901.04442 [hep-ph]].

\bibitem{Kallweit:2017khh}
S.~Kallweit, J.~M.~Lindert, S.~Pozzorini and M.~Sch\"onherr,
JHEP \textbf{11} (2017), 120
[arXiv:1705.00598 [hep-ph]].


\bibitem{Brauer:2020kfv}
S.~Br\"auer, A.~Denner, M.~Pellen, M.~Sch\"onherr and S.~Schumann,
JHEP \textbf{10} (2020), 159
[arXiv:2005.12128 [hep-ph]].

\bibitem{Granata:2017iod}
F.~Granata, J.~M.~Lindert, C.~Oleari and S.~Pozzorini,
JHEP \textbf{09} (2017), 012
[arXiv:1706.03522 [hep-ph]].

\bibitem{Chiesa:2019ulk}
M.~Chiesa, A.~Denner, J.~N.~Lang and M.~Pellen,
Eur. Phys. J. C \textbf{79} (2019) no.9, 788
[arXiv:1906.01863 [hep-ph]].

\bibitem{Chiesa:2020ttl}
M.~Chiesa, C.~Oleari and E.~Re,
Eur. Phys. J. C \textbf{80} (2020) no.9, 849
[arXiv:2005.12146 [hep-ph]].

\bibitem{Barze:2012tt}
L.~Barze, G.~Montagna, P.~Nason, O.~Nicrosini and F.~Piccinini,
JHEP \textbf{04} (2012), 037
[arXiv:1202.0465 [hep-ph]].

\bibitem{Bernaciak:2012hj}
C.~Bernaciak and D.~Wackeroth,
Phys. Rev. D \textbf{85} (2012), 093003
[arXiv:1201.4804 [hep-ph]].

\bibitem{Barze:2013fru}
L.~Barze, G.~Montagna, P.~Nason, O.~Nicrosini, F.~Piccinini and A.~Vicini,
Eur. Phys. J. C \textbf{73} (2013) no.6, 2474
[arXiv:1302.4606 [hep-ph]].

\bibitem{CarloniCalame:2016ouw}
C.~M.~Carloni Calame, M.~Chiesa, H.~Martinez, G.~Montagna, O.~Nicrosini, F.~Piccinini and A.~Vicini,
Phys. Rev. D \textbf{96} (2017) no.9, 093005
doi:10.1103/PhysRevD.96.093005
[arXiv:1612.02841 [hep-ph]].

\bibitem{Muck:2016pko}
A.~M\"uck and L.~Oymanns,
JHEP \textbf{05} (2017), 090
[arXiv:1612.04292 [hep-ph]].

\bibitem{Chiesa:2019nqb}
M.~Chiesa, F.~Piccinini and A.~Vicini,
Phys. Rev. D \textbf{100} (2019) no.7, 071302
[arXiv:1906.11569 [hep-ph]].


\bibitem{Bothmann:2020sxm}
E.~Bothmann and D.~Napoletano,
Eur. Phys. J. C \textbf{80} (2020) no.11, 1024
[arXiv:2006.14635 [hep-ph]].

\bibitem{Denner:2000jv}
A.~Denner and S.~Pozzorini,
Eur. Phys. J. C \textbf{18} (2001), 461-480
[arXiv:hep-ph/0010201 [hep-ph]].

\bibitem{hybrid_psr_21}
E. Bothmann,
talk at ``Parton Showers and Resummation'' 2021 (PSR 2021), https://indico.cern.ch/event/1018828

\bibitem{taming}
``Taming the accuracy of event generators'', https://indico.cern.ch/event/999271/


\end{thebibliography}
\end{document}